\documentclass[aps,prl,twocolumn,superscriptaddress]{revtex4}%
\usepackage{graphicx}
\usepackage{dcolumn}
\usepackage{amsmath}
\usepackage{amssymb}
\usepackage{indentfirst}
\usepackage{bm}
\usepackage{multirow}
\usepackage{graphicx}
\usepackage{hyperref}
\usepackage{ifpdf}
\begin{document}

\title{Nonadiabatic Geometric Quantum Computation in Non-Hermitian Systems}
\author{Tian-Xiang Hou}
\affiliation{
School of Physics and Electronic Information, Yanan University, Yanan 716000, China}

\author{Wei Li} \email{liweixici@126.com}
\affiliation{
School of Physics and Electronic Information, Yanan University, Yanan 716000, China}

\date{\today}
\begin{abstract}
Nonadiabatic geometric quantum computation (NGQC) has emerged as an excellent proposal for achieving fast and robust quantum control against control errors. However, previous NGQC protocols could not be strongly resilient against the noise from decay of bare states in a realistic system, which can be equivalently described by a non-Hermitian Hamiltonian. Here, we show how to perform NGQC in non-Hermitian quantum systems. By utilizing a novel geometric phase generated by non-unitary evolution of the system, a universal set of geometric gates can be realized with a high fidelity. Moreover, we demonstrate that the nonadiabatic process does not lead to
the loss of fidelity from decay.
\end{abstract}
\maketitle
Quantum computation is regarded as a more powerful and effective tool than classical computation in tackling some issues, including quantum simulation \cite{simulation1}, searching unsorted data \cite{data}, and prime factorizations of large integers \cite{integers}. To achieve these tasks, it is necessary to construct quantum gates that are fast and accurately controllable. Geometric quantum computation (GQC) is believed as an excellent proposal for quantum control due to its reliance on global geometrical features of evolution trajectories, making it robust against specific errors \cite{local1,local2,local3}.

Early schemes of GQC \cite{GQC1,GQC2,GQC3}, based on adiabatic geometric phase specified by either a real number, i.e., Berry phase \cite{Berry}, or a matrix \cite{matrix}, require systems to evolve slowly in order to operate quantum gates. Nevertheless, in the adiabatic process, quantum gates are prone to sensitivity from environment-induced decoherence. Later, various approaches to nonadiabatic geometric quantum computation (NGQC) were proposed to tackle this issue. For example, the family of nonadiabatic Abelian GQC approaches employs Aharonov-Anandan (AA) phase \cite{AA} induced by single loop \cite{single1,single2,single3,single4,single5} or multiple loops \cite{multiple1,multiple2,multiple3} evolution of a two-level system to perform the desired gate. The schemes involving nonadiabatic holonomic quantum computation (NHQC) \cite{holonomic1,holonomic2}, composite NHQC \cite{composite}, single-loop multiple-pulse NHQC \cite{NHQC}, and NHQC+ \cite{NHQC+} can realize a universal set of one-qubit gates by utilizing a three-level system, of which two lower levels encoding a qubit are coupled to the excited state by elaborate pulses. Since NGQC schemes \cite{single1,single2,single3,single4,single5,multiple1,multiple2,multiple3,holonomic1,holonomic2,composite,NHQC,NHQC+,schemes1,schemes2,schemes3} possess the advantages of quantum control for fast and robust evolution, which have been experimentally realized in the systems of superconducting circuits \cite{circuits1,circuits2,circuits3}, nuclear magnetic resonance \cite{resonance1,resonance2}, trapped ions \cite{ions}, and nitrogen-vacancy (NV) center in diamond \cite{diamond1,diamond2,diamond3}.

To realize large-scale NGQC, quantum gates should be strongly resilient against various noises induced by surrounding environment. One of the significant sources of noise originates from the decay of bare states in a realistic system, which can be described equivalently by a non-Hermitian Hamiltonian. However, previous NGQC protocols are limited to Hermitian quantum systems, which prompts us to explore a new NGQC scheme in non-Hermitian quantum systems, aiming to achieve the following goals: (1) possessing real purely geometric phases; (2) robustness against the control errors; (3) no loss of fidelity of quantum gates due to decay.

In this letter, we demonstrate that the exactly nonadiabatic periodic evolution of a non-Hermitian system, driven by elaborate complex pulses, can induce a novel geometric phase, which is just the real part of complex A-A phase and corresponds to the real part of complex solid angle in complex parameter space. Based on the geometric phase, a universal set of geometric gates can be realized with a high fidelity against certain systematic error and control errors. In addition, the present scheme shows that the non-unitary process does not result in the loss of fidelity.

\textit{Construction of Hamiltonian for non-Hermitian quantum systems.}
In the frame of biorthonormal set, we start by presenting the general form of the time evolution operator for a N-dimensional non-Hermitian quantum system, and then design Hamiltonian $H(t)$ to be used for realization of NGQC. Specially, one introduces a set $\{|\psi_{m}(t)\rangle, |\tilde{\psi}_{m}(t)\rangle\}_{m=1}^{N}$ with the biorthonormal
condition $\langle\tilde{\psi}_{m}(t)|\psi_{n}(t)\rangle=\langle\psi_{m}(t)|\tilde{\psi}_{n}(t)\rangle=\delta_{mn}$ and the completeness relation
$\sum_{m}|\tilde{\psi}_{m}(t)\rangle \langle\psi_{m}(t)|=\sum_{m}|\psi_{m}(t)\rangle \langle\tilde{\psi}_{m}(t)|=I$ at every time moment $t$. Here the states $|\psi_{m}(t)\rangle$ and $|\tilde{\psi}_{m}(t)\rangle$ satisfy the Schr\"{o}dinger equation $i\partial_{t}|\psi_{m}(t)\rangle=H(t)|\psi_{m}(t)\rangle$ and its adjoint equation $i\partial_{t}|\tilde{\psi}_{m}(t)\rangle=H^{\dag}(t)|\tilde{\psi}_{m}(t)\rangle$, respectively. The time evolution operator, which precisely illustrates the exact evolution from the initial state $|\psi_{m}(t_0)\rangle$ along the specific path described by state $|\psi_{m}(t)\rangle$, can be then expressed as $U(t,t_0)=\sum_{m}|\psi_{m}(t)\rangle \langle\tilde{\psi}_{m}(t_0)|$.

To further construct geometric phase, we employ auxiliary complete biorthonormal set $\{|\phi_{m}(t)\rangle, |\tilde{\phi}_{m}(t)\rangle\}_{m=1}^{N}$ with the vectors defined by
\begin{eqnarray}
|\phi_{m}(t)\rangle&=&e^{-i\alpha_m(t)}|\psi_{m}(t)\rangle, \nonumber \\
|\tilde{\phi}_{m}(t)\rangle&=&e^{-i\alpha_{m}^{*}(t)}|\tilde{\psi}_{m}(t)\rangle. \label{relation}
\end{eqnarray}
These vectors satisfy the periodic boundary conditions
\begin{eqnarray}
\begin{aligned}
|\phi_{m}(t_0)\rangle&=|\phi_{m}(t_f)\rangle, &
|\tilde{\phi}_{m}(t_0)\rangle&=|\tilde{\phi}_{m}(t_f)\rangle. \label{boundary}
\end{aligned}
\end{eqnarray}
The evolution operator can then be recast as
\begin{equation}
U(t,t_0)=\sum_{m}e^{i\alpha_m(t)}|\phi_{m}(t)\rangle \langle\tilde{\phi}_{m}(t_0)|. \label{U}
\end{equation}
By making use of $i\partial_{t}U(t,t_0)=H(t)U(t,t_0)$, the exact expression of Hamiltonian $H(t)$ can be worked out to be
\begin{equation}
H(t)=\sum_{m}i|\dot{\phi}_{m}(t)\rangle \langle\tilde{\phi}_{m}(t)|-\dot{\alpha}_{m}(t)|\phi_{m}(t)\rangle\langle\tilde{\phi}_{m}(t)|,    \label{H}
\end{equation}
with the phase factor given by $\alpha_m(t)=\int_{t_0}^{t}\langle\tilde{\phi}_{m}(t')|i\partial_{t'}-H(t')|\phi_{m}(t')\rangle dt'$. In general, the phase factor is a time-dependent complex function. However, for the realization of geometric gates, one expects it is real and purely geometric, that is,
\begin{equation}
\alpha_{m}(t)=\text{Re}\int_{t_0}^{t}\langle\tilde{\phi}_{m}(t')|i\partial_{t'}|\phi_{m}(t')\rangle dt'.  \label{phase}
\end{equation}
It should be noted that, even though the evolution is often non-unitary, it is desired that the operator $U(t,t_0)$ specified by Eq. (\ref{U}) should be unitary at time $t=t_{0,f}$, which implies $\{|\phi_{m}(t_0)\rangle\}$ and $\{|\tilde{\phi}_{m}(t_0)\rangle\}$ are two complete sets,
\begin{eqnarray}
\begin{aligned}
\langle\phi_{m}(t_0)|\phi_{n}(t_0)\rangle&=\delta_{mn}, &
\sum_{m}|\phi_{m}(t_0)\rangle \langle\phi_{m}(t_0)|&=I, \nonumber
\end{aligned}
\\
\begin{aligned}
\langle\tilde{\phi}_{m}(t_0)|\tilde{\phi}_{n}(t_0)\rangle&=\delta_{mn}, &
\sum_{m}|\tilde{\phi}_{m}(t_0)\rangle \langle\tilde{\phi}_{m}(t_0)|&=I. \label{boundary1}
\end{aligned}
\end{eqnarray}
Consequently, the time evolution operator at the final time $t=t_f$ reads
\begin{equation}
U(t_f,t_0)=\sum_{m}e^{i\alpha_{m}(t_f)}|\phi_{m}(t_0)\rangle \langle\tilde{\phi}_{m}(t_0)|. \label{unitary1}
\end{equation}

In the following section, we will concentrate on the non-Hermitian two-level system and provide the explicit form of Hamiltonian Eq. (\ref{H}) by selecting the base vectors of the set $\{|\phi_{m}(t)\rangle, |\tilde{\phi}_{m}(t)\rangle\}$ that satisfy the conditions in Eq. (\ref{boundary1}). Additionally, we will investigate the geometric feature of the geometric phase $\alpha_{m}(t_f)$ specified by Eq. (\ref{phase}) and demonstrate that the dynamics of the system can be utilized to implement arbitrary one-qubit geometric gates.

\textit{Construction of arbitrary one-qubit gate.}
Consider a driven non-Hermitian two-level system, with lower and upper bare states labelled by $|+\rangle=(1,0)^{T}$ and $|-\rangle=(0,1)^{T}$, respectively. We expand upon the concept of the ``Bloch sphere" to a non-Hermitian two-level system, called ``complex Bloch sphere" \cite{Complex}.
The base vectors take the following form
\begin{eqnarray}
|\phi_{\pm}(t)\rangle&=&\cos{\frac{\tilde{\theta}(t)}{2}}|\pm\rangle\pm\sin{\frac{\tilde{\theta}(t)}{2}}e^{\pm i\tilde{\varphi}(t)}|\mp\rangle, \nonumber \\
|\tilde{\phi}_{\pm}(t)\rangle&=&\cos{\frac{\tilde{\theta}^{*}(t)}{2}}|\pm\rangle\pm\sin{\frac{\tilde{\theta}^{*}(t)}{2}}e^{\pm i\tilde{\varphi}^{*}(t)}|\mp\rangle,  \label{states}
\end{eqnarray}
in which $\tilde{\theta}(t)$ and $\tilde{\varphi}(t)$ are usually time-dependent complex functions. However, following the boundary conditions specified by Eqs. (\ref{boundary}) and (\ref{boundary1}), they should be real at time $t=t_{0,f}$. The real geometric phase in Eq. (\ref{phase}) can be worked out to be
\begin{equation}
\alpha_{\pm}(t) =\mp \text{Re}\int_{t_0}^{t}(\dot{\tilde{\varphi}}\sin^{2}\frac{\tilde{\theta}}{2})dt'.    \label{geometric}
\end{equation}
Furthermore, plugging Eqs. (\ref{states}) and (\ref{geometric}) into Eq. (\ref{H}), one has the Hamiltonian,
\begin{equation}
H(t)=\frac{1}{2}[\Omega_{x}(t)\sigma_{x}+\Omega_{y}(t)\sigma_{y}+\Omega_{z}(t)\sigma_{z}], \label{Hamiltonian}
\end{equation}
where the explicit expression of complex pulse parameters are given by
\begin{eqnarray}
\Omega _x(t)&=&\tilde{f}(t)\sin{\tilde{\theta}}(t)\cos{\tilde{\varphi}}(t)-\dot{\tilde{\theta}}(t)\sin{\tilde{\varphi}}(t),   \nonumber \\
\Omega _y(t)&=&\tilde{f}(t)\sin{\tilde{\theta}}(t)\sin{\tilde{\varphi}}(t)+\dot{\tilde{\theta}}(t)\cos{\tilde{\varphi}}(t),   \label{externals} \\
\Omega _z(t)&=&\tilde{f}(t)\cos{\tilde{\theta}}(t)+\dot{\tilde{\varphi}}(t),  \nonumber
\end{eqnarray}
with $\tilde{f}(t)=2\text{Re}(\dot{\tilde{\varphi}}\sin^{2}\frac{\tilde{\theta}}{2})-\dot{\tilde{\varphi}}$. Here $\sigma_{x,y,z}$ denote the Pauli operator acting on two-dimensional Hilbert space.

We now demonstrate that the Hamiltonian can be used to describe a general non-Hermitian two-level system, such as Rydberg atom, with energy gap $\omega_{0}$. The system is driven by a complex pulse with two components: $\bm{E}(t)=\bm{E}_{1}(t)\cos[\int_{t_0}^{t} \omega(t')dt'+\phi_{1}(t)]+i\bm{E}_{2}(t)\cos[\int_{t_0}^{t} \omega(t')dt'+\phi_{2}(t)]$. Here $\bm{E}_{i}(t)$ and $\phi_{i}(t)(i=1,2)$ represent the time-dependent amplitude and phase of the ith component of the pulse, and $\omega(t)$ denotes the pulse frequency. Such complex pulse could be realized in optical systems \cite{optical1,optical2}. The decay rates from the states $|+\rangle$ and $|-\rangle$ are denoted by $\gamma_{+}(t)$ and $\gamma_{-}(t)$, respectively. Under the rotating wave approximation, the Hamiltonian in interaction picture can be expressed as
\begin{equation}
\begin{split}
H(t)&=\frac{1}{2}\{[\Omega_{1}(t)\cos\phi_{1}(t)+i\Omega_{2}(t)\cos\phi_{2}(t)]\sigma_{x}\\
&+[\Omega_{1}(t)\sin\phi_{1}(t)+i\Omega_{2}(t)\sin\phi_{2}(t)]\sigma_{y} \\
&+[\Delta(t)-i\delta(t)]\sigma_{z}\},
\end{split}
\label{new}
\end{equation}
where $\Omega_{1,2}(t)=-\langle+|\bm{d}\cdot\bm{E}_{1,2}(t)|-\rangle$ are the Rabi frequencies with $\bm{d}$ being the atomic transition dipole moment, and $\Delta(t)=\omega(t)-\omega_{0}$ denotes the detuning from resonance. $\delta(t)=(\gamma_{+}(t)-\gamma_{-}(t))/2$ represents the difference between the decay rates \cite{decay}. By comparing the expression of Hamiltonian in Eq. (\ref{new}) with Eq. (\ref{Hamiltonian}), one recognizes that the Rabi frequencies correspond to
$\Omega_{1}(t)=\sqrt{[\text{Re}\Omega_{x}(t)]^{2}+[\text{Re}\Omega_{y}(t)]^{2}}$ and $\Omega_{2}(t)=\sqrt{[\text{Im}\Omega_{x}(t)]^{2}+[\text{Im}\Omega_{y}(t)]^{2}}$.
The real and imaginary parts of pulse parameter $\Omega_{z}(t)$ are associated with $\Delta(t)$ and $\delta(t)$, i.e., $\Delta(t)=\text{Re}[\Omega_z(t)]$ and $\delta(t)=-\text{Im}[\Omega_z(t)]$. The phases of the pulse are characterized by $\phi_{1}(t)=\arccos\frac{\text{Re}[\Omega_{x}(t)]}{\Omega_{1}(t)}$ and $\phi_{2}(t)=\arccos\frac{\text{Im}[\Omega_{x}(t)]}{\Omega_{2}(t)}$. Considering that $\delta(t)$ is a function of $\tilde{\theta}(t)$ and $\tilde{\varphi}(t)$, all the pulse parameters in Eq. (\ref{externals}) can be related to $\delta(t)$. It implies that there is no need to avoid the impact of decay which is solely dependent on the intrinsic properties of the system and can be measured experimentally. On the contrary, the time-dependent decay rates can be matched by adjusting the external field $\bm{E}(t)$ in a way that does not affect the fidelity of the quantum gates described in Eq. (\ref{unitary1}).

Let us move to manifest the geometric feature of the geometric phase $\alpha_{m}(t_f)$ specified by Eq. (\ref{phase}) in a non-Hermitian two-level system. In complex parameter space characterized by the complex vector $\tilde{\bm{n}}(\tilde{\theta},\tilde{\varphi})$, the states $|\phi_{+}(t)\rangle$ and $\langle\tilde{\phi}_{+}(t)|$, as specified by Eqs. (\ref{states}), will move along the same continuous and smooth closed curve $\tilde{C}$ during the periodic evolution over $t\in [t_0, t_f]$. This allows us to calculate $\alpha_{\pm}(t_f)$ as a curve integration along the path $\tilde{C}$,
$\alpha_{\pm}(t_f)=\pm \text{Re}\oint_{\tilde{C}} \bm{\tilde{A}}(\tilde{\bm{n}})\cdot d\tilde{\bm{n}}$ with the integrand $\bm{\tilde{A}}(\tilde{\bm{n}})=i\langle\tilde{\phi}_{+}(\tilde{\bm{n}})|\partial_{\tilde{\bm{n}}}|\phi_{+}(\tilde{\bm{n}})\rangle$ defined as complex A-A connection (see Supplemental Material \cite{SM}). The direction of curve $\tilde{C}$ has been assumed to be counterclockwise, as determined by the right-hand rule for the outward normal of the surface. To further apply the Stock theorem, the line integration can be recast as a surface integral over the enclosed surface $\tilde{\Sigma}$ bounded by closed path $\tilde{C}$,
$\alpha_{\pm}(t_f)=\pm \text{Re}\iint_{\tilde{\Sigma}} \partial_{\tilde{\bm{n}}}\times\bm{\tilde{A}}(\tilde{\bm{n}})\cdot d\tilde{\bm{S}}$. We finally obtain $\alpha_{\pm}(t_f)=\mp \text{Re}(\tilde{\Omega}(\tilde{C})/2)$ with $\tilde{\Omega}(\tilde{C})=\iint_{\tilde{\Sigma}}\sin\tilde{\theta}d\tilde{\theta}d\tilde{\varphi}$ as the complex solid angle swept by the closed evolution. Remarkably, the geometric phase is dependent on the complex solid angle, while remaining independent of changing rate of the complex parameters or speed of evolving states. It indicates that the quantum gates realized are immune to the control errors only affecting dynamical process but no changing the solid angle. Moreover, for a particular case with only $\tilde{\varphi}(t)$ as a complex angle, the geometric phase consists with that of the conventional A-A phase in Hermitian quantum systems \cite{space}.

Next, we show that the dynamics of the above two-level system can realize an arbitrary one-qubit gate. Specifically, the system will undergo a cyclic evolution along a closed path $\tilde{C}$ with $\bm{\tilde{n}}(t_f)=\bm{\tilde{n}}(t_0)$, based on the states in Eqs. (\ref{states}), the evolution operator specified by Eq. (\ref{unitary1}) is further formulated as
\begin{equation}
U_{\theta,\varphi,\alpha}(\tilde{C})=e^{-i\alpha\bm{n}\cdot\bm{\sigma}},  \label{gate}
\end{equation}
where $\bm{n}=(\sin\theta\cos\varphi,\sin\theta\sin\varphi,\cos\theta)$ is a real unit vector with $\theta=\tilde{\theta}(t_0)$ and $\varphi=\tilde{\varphi}(t_0)$, and the coefficient is given by $\alpha=\alpha_{-}(t_f)$.
Since both $\bm{n}$ and $\alpha$ can take any value, $U(\tilde{C})$ describes an arbitrary Abelian geometric one-qubit gate operation with real rotation axis $\bm{n}$ and rotation angle $2\alpha$. It should be emphasized that the unitary quantum gates guarantee the nonadiabatic process does not lead to the loss of fidelity from decay relative to that in the previous GQC approaches \cite{single1,single2,single3,single4,single5,multiple1,multiple2,multiple3,holonomic1,holonomic2,composite,NHQC,NHQC+,schemes1,schemes2,schemes3}.

\textit{The effects of the deviation in polar angle.}
The nonadiabatic process of a non-Hermitian quantum system can be depicted by the complex angles $\{\tilde{\theta}(t), \tilde{\varphi}(t)\}$. The deviation of the angles not only results in the inaccurate control of the pulse but also causes a reduction in fidelity of one-qubit gate. Here we only explore the effects of the polar angle deviation $\delta{\tilde{\varphi}}=\tilde{\varphi}'(t)-\tilde{\varphi}(t)$ on the parameters $\{\Omega_{1}(t),\Omega_{2}(t),\phi_{1}(t),\phi_{2}(t),\Delta(t)\}$ \cite{azimuthal}. We assume that $\tilde{\varphi}(t)$ fluctuates slowly as compared to the gate operation such that $\delta{\tilde{\varphi}}=\delta\varphi_{1}+i\delta\varphi_{2}$ is a time-independent small quantity, which guarantees the geometric phase specified by Eq. (\ref{geometric}) does not change with this deviation. By plugging $\tilde{\varphi}'(t)$ into Eq. (\ref{externals}), one has $\Omega_{1}(t)\Omega_{2}(t)\cos(\phi_{1}(t)-\phi_{2}(t))=\Omega_{1}'(t)\Omega_{2}'(t)\cos(\phi_{1}'(t)-\phi_{2}'(t))$, which shows that the imperfect control Rabi frequencies and phases are not independent. Nevertheless, the detuning is immune to the deviation of the polar angle due to $\Delta'(t)=\Delta(t)$.

To proceed, according to Eq. (\ref{gate}), the deviation $\delta{\tilde{\varphi}}$ leads to the imperfect evolution characterized by the operation $V=e^{-i\alpha\bm{n}'\cdot\bm{\sigma}}$
with the erroneous rotation axis $\bm{n}'(\theta,\varphi+\delta{\tilde{\varphi}})$. To reveal the influence of this deviation on Abelian geometric one-qubit gate, we define the fidelity $F=\frac{|\text{Tr}(VU^{\dagger})|}{|\text{Tr}(VV^{\dagger})|}$, in which $U=e^{-i\alpha\bm{n}\cdot\bm{\sigma}}$ describes the ideal evolution. Here the added denominator ensures the fidelity $F\in[0,1]$.
We further consider a special case with only the derivation of the real part of polar angle involved, i.e., $\delta\varphi_{2}=0$. Following the expression in Eq. (\ref{externals}), one has the Rabi frequencies $\Omega_{1,2}'(t)=\Omega_{1,2}(t)$. The systematic error from inaccurate control of pulse phases could be reflected by $\delta\varphi_{1}$, that is, $\delta\phi_{1,2}=\delta\varphi_{1}$. The fidelity in this case can reach $F=1-\frac{1}{2}\sin^{2}\alpha\sin^{2}\theta\delta\varphi_{1}^{2}$, which is the same as that in \cite{fidelity}. Compared with dynamical gates $F_{d}=1-\frac{1}{2}\sin^{2}\theta_{0}\delta\phi^{2}$ \cite{dynamical} and nonadiabatic non-Abelian geometric gates $F_{g}=1-\frac{1}{2}\sin^{2}\alpha\delta\phi^{2}$ \cite{holonomic1}, it is evident that the geometric gates in the present scheme demonstrate greater robustness against the systematic error from the phases even if the nonadiabatic process is non-unitary.

\textit{Realization of controlled-phase gate.}
To realize NGQC, a controlled-phase gate, in addition to arbitrary one-qubit gates, is used to construct the universal set. We will demonstrate how to implement controlled-phase gate in a non-Hermitian system composed of two identical neutral atoms with hyperfine states $|g\rangle_{j}$ and $|e\rangle_{j}$ and Stark eigenstate $|r\rangle_{j}$ ($j=1,2$ label the atoms) \cite{system}. Specially, we utilize a two component of complex pulse $\bm{E}(t)=\bm{E}_{1}(t)\cos[\int_{t_0}^{t} \omega(t')dt'+\phi_{1}(t)]+i\bm{E}_{2}(t)\cos[\int_{t_0}^{t} \omega(t')dt'+\phi_{2}(t)]$ to drive the two atomic transitions $|e\rangle_{j}\leftrightarrow|r\rangle_{j}$, while no pulses affect the states $|g\rangle_{j}$. The dipole-dipole interaction between the atoms leads to an energy shift, denoted by $u$, when they populate the state $|r\rangle_{1}|r\rangle_{2}$. Under the rotating wave approximation, the system is described by the Hamiltonian
\begin{equation}
\begin{split}
H(t)&=u|r\rangle_{1}\langle r|\otimes |r\rangle_{2}\langle r|\\
&+\frac{1}{2}\sum_{j=1,2}\Big[\Omega_{1}(t)e^{-i\phi_{1}(t)}|e\rangle_{j}\langle r|+i\Omega_{2}(t)e^{i\phi_{2}(t)}|r\rangle_{j}\langle e|\\
&+[\Delta(t)-i\delta(t)](|e\rangle_{j}\langle e|-|r\rangle_{j}\langle r|)\Big],
\end{split}
\label{Hamiltonian1}
\end{equation}
where $\Omega_{1,2}(t)=-\langle+|\bm{d}\cdot\bm{E}_{1,2}(t)|-\rangle$ are Rabi frequencies, $\Delta(t)$ is the detuning of the pulse, and $\delta(t)= (\gamma_{r}(t)-\gamma_{e}(t))/2$ accounts the difference between the decay rates. Drawing upon similarities with \textit{model A} discussed in Ref. \cite{system}, we assume that $u\ll\min{(\Omega_{1,2}(t))}$ to ensure the coupling between pulses and atoms is much larger than the dipole-dipole interaction between atoms, the controlled-Z gate can be performed as the following steps: (1) apply $iX$ gate to the j-th atom if it is initially in the state $|e\rangle_{j}$; (2) no operations are applied to the atoms during the subsequent time interval $\Delta t=\Theta/u$; (3) apply $-iX$ gate to the same atom that was operated in step (1). According to Eq. (\ref{gate}), $iX$ and $-iX$ gates can be realized by setting parameters $(\theta,\varphi,\alpha)=(\frac{\pi}{2},\pi,\frac{\pi}{2})$ and $(\theta,\varphi,\alpha)=(\frac{\pi}{2},0,\frac{\pi}{2})$, respectively. It is worth noting that the state $|gg\rangle$ remains unaffected since no operations are applied to it. When the initial state is set as $|eg\rangle$ or $|ge\rangle$, the gate sequence consisting of $iX$ and $-iX$ acts on the individual atomic state $|e\rangle_{j}$, resulting in $|eg\rangle\rightarrow |eg\rangle$ and $|ge\rangle\rightarrow |ge\rangle$. If the system initially populates the state $|ee\rangle$, the first gate $iX\otimes iX$ will transfer $|ee\rangle$ to the state $-|rr\rangle$, which then accumulates a phase $\Theta=u\Delta t$ during the interval. The second gate $(-iX)\otimes (-iX)$ will subsequently bring the state back to $e^{i\Theta}|ee\rangle$. By labeling $|g\rangle_{j}=|0\rangle_{j}$ and $|e\rangle_{j}=|1\rangle_{j}$, the controlled-phase gate $U=(1,1,1,e^{i\Theta})$ is then realized in the computing basis $\{|00\rangle,|01\rangle,|10\rangle,|11\rangle\}$.

\textit{Conclusion.} In this letter, we have put forward a scheme to perform NGQC in non-Hermitian quantum systems. By introducing a complete biorthonormal set, we can design the desired time-dependent Hamiltonian. The purely geometric phase, induced by non-unitary and periodic evolution of a two-level system, is determined as the real part of complex A-A phase. Importantly, in parameter space, this geometric phase only depends on the complex solid angle enclosed by a loop path but is independent of changing rate of the parameters or speed of evolving states. Therefore, geometric gates realized are immune to the control errors only affecting dynamical process but no changing the solid angle. Moreover, we reveal that arbitrary one-qubit gate in the present scheme is more robust against systematic error from the deviation of pulse phases than both dynamical gates and nonadiabatic non-Abelian geometric gates.

To demonstrate the application of the scheme, we utilize two-component complex pulses to manipulate neutral atoms with long coherence times, allowing for the implementation of arbitrary single-qubit gates and two-qubit controlled-phase gates. We have demonstrated that the driven complex pulse can be adjusted with the time-dependent decay rates of a quantum system, such that the nonadiabatic process does not lead to the loss of fidelity from decay relative to that in the previous GQC approaches.
\bigskip

\acknowledgments
The authors thank Zhang-qi Yin, Li-xiang Cen and Jian-wei Xu for fruitful discussion. This work was supported by the Scientific Research Program of Education Department of Shaanxi Provincial Government (Grant No. 22JK0617).

\section{geometric phase and A-A connection}
The geometric phase of a non-Hermitian two-level system is given by
\begin{equation}
\alpha_{\pm}(t_f)=\text{Re}\int_{t_0}^{t_f}\langle\tilde{\phi}_{\pm}(t')|i\partial_{t'}|\phi_{\pm}(t')\rangle dt'.  \label{phase11}
\end{equation}
In the complex parameter space, we assume that, during the periodic evolution over $t\in [t_0, t_f]$, the states $|\phi_{+}(t)\rangle$ and $|\phi_{-}(t)\rangle$ will move along continuous and smooth closed curve $\tilde{C}$ and $\tilde{C'}$, respectively. Following the Eq. (8) in the main text, it is seen that the points of curve $\tilde{C}$ and $\tilde{C'}$ are corresponding to complex vectors $\tilde{\bm{n}}(\tilde{\theta},\tilde{\varphi})$ and $\tilde{\bm{n}}'(\tilde{\theta},\pi-\tilde{\varphi})$. Therefore, we change the variable $t$ into $\tilde{\bm{n}}$ and $\tilde{\bm{n}}'$ and obtain that
\begin{eqnarray}
\alpha_{+}(t_f)&=&i\text{Re}\oint_{\tilde{C}} \langle\tilde{\phi}_{+}(\tilde{\bm{n}})|\partial_{\tilde{\bm{n}}}|\phi_{+}(\tilde{\bm{n}})\rangle\cdot d\tilde{\bm{n}}   \nonumber \\
&=&-i\text{Re}\oint_{\tilde{C'}} \langle\tilde{\phi}_{-}(\tilde{\bm{n}}')|\partial_{\tilde{\bm{n}}'}|\phi_{-}(\tilde{\bm{n}}')\rangle\cdot d\tilde{\bm{n}}'.   \label{phase21}
\end{eqnarray}
By introducing $\bm{\tilde{A}}(\tilde{n})=i\langle\tilde{\phi}_{+}(\tilde{\bm{n}})|\partial_{\tilde{\bm{n}}}|\phi_{+}(\tilde{\bm{n}})\rangle$, the geometric phase in Eq. (\ref{phase11}) can be recast as
\begin{equation}
\alpha_{\pm}(t_f)=\pm \text{Re}\oint_{\tilde{C}} \bm{\tilde{A}}(\tilde{\bm{n}})\cdot d\tilde{\bm{n}}. \label{phase}
\end{equation}
We now demonstrate that the integrand $\bm{\tilde{A}}(\tilde{\bm{n}})$, referred to as the complex A-A connection, represents a well-defined connection. Since the state $|\phi_{+}(\tilde{\bm{n}})\rangle$ is normalized in the framework of biorthonormal set, the connection can be calculated as
\begin{eqnarray}
\bm{\tilde{A}}(\tilde{n})&=&i\langle\tilde{\phi}_{+}(\tilde{\theta},\tilde{\varphi})|(\bm{e}_{\tilde{\theta}}\partial_{\tilde{\theta}}+\frac{\bm{e}_{\tilde{\varphi}}}{\sin{\tilde{\theta}}}\partial_{\tilde{\varphi}})|\phi_{+}(\tilde{\theta},\tilde{\varphi})\rangle,   \nonumber \\
&=&-\frac{1-\cos{\tilde{\theta}}}{\sin{\tilde{\theta}}}\bm{e}_{\tilde{\varphi}},   \label{externals22}
\end{eqnarray}
which is continuous except at the singularities located at $\tilde{\theta}=0$ (north pole) and $\tilde{\theta}=\pi$ (south pole). We further define the complex A-A curvature as
\begin{equation}
\bm{B}(\tilde{\bm{n}})=\partial_{\tilde{\bm{n}}}\times\bm{\tilde{A}}(\tilde{\bm{n}}). \label{B}
\end{equation}
By substituting Eq. (\ref{externals22}) into Eq. (\ref{B}), one obtains $\bm{B}(\tilde{\bm{n}})=-\frac{1}{2}\bm{e}_{\tilde{r}}$, where $\bm{e}_{\tilde{r}}$ indicates the direction of the outward normal of the surface. It is notable that the singularities in the connection do not induce divergency in the curvature. Consequently, the so called complex A-A phase, which can be represented as a surface integral over the enclosed surface $\tilde{\Sigma}$ bounded by closed
path $\tilde{C}$
\begin{equation}
\alpha_{\pm}^{AA}=\iint_{\tilde{\Sigma}} \bm{B}(\tilde{\bm{n}})\cdot d\tilde{\bm{S}}, \label{phase3}
\end{equation}
remains unaffected even if the path of quantum state transverses these singularities. Moreover, due to the constant norm of $\bm{B}(\tilde{\bm{n}})$, we can directly obtain $\partial_{\tilde{\bm{n}}}\cdot\bm{B}(\tilde{\bm{n}})=0$. Thus, in parameter space the complex A-A curvature, complex A-A connection, and complex A-A phase can be associated with the notions of `magnetic field strength', `magnetic vector potential', and  `magnetic flux', respectively. Importantly, if the states undergo phase transformations
\begin{eqnarray}
|\phi_{+}(\tilde{\bm{n}})\rangle&\rightarrow& e^{i\chi(\tilde{\bm{n}})}|\phi_{+}(\tilde{\bm{n}})\rangle, \nonumber \\
|\tilde{\phi}_{+}(\tilde{\bm{n}}^{*})\rangle&\rightarrow& e^{i\chi^{*}(\tilde{\bm{n}}^{*})}|\tilde{\phi}_{+}(\tilde{\bm{n}}^{*})\rangle,   \nonumber   \label{transformation1}
\end{eqnarray}
which is equivalent to the `magnetic vector potential' undergoing a gauge transformation
\begin{equation}
\bm{\tilde{A}}(\tilde{\bm{n}})\rightarrow\bm{\tilde{A}}(\tilde{\bm{n}})-\partial_{\tilde{\bm{n}}}\chi(\tilde{\bm{n}}), \label{transformation2}
\end{equation}
then both the `magnetic field strength' $\bm{B}(\tilde{\bm{n}})$ and the real part of `complex magnetic flux' $\alpha_{\pm}(t_f)=\text{Re}(\alpha_{\pm}^{AA})$ are invariant under the aforementioned gauge transformation.


\begin{thebibliography}{99}
\bibitem{simulation1} F. Pan and P. Zhang, Simulation of Quantum Circuits Using the Big-Batch Tensor Network Method, Phys. Rev. Lett. {\bf 128}, 030501 (2022).
\bibitem{data} L. K. Grover, Quantum Mechanics Helps in Searching for a Needle in a Haystack, Phys. Rev. Lett. {\bf 79}, 325 (1997).
\bibitem{integers} Shor and W. Peter, Polynomial-Time Algorithms for Prime Factorization and Discrete Logarithms on a Quantum Computer, SIAM J. Comput. {\bf 26}, 1484 (1997).
\bibitem{local1} S. L. Zhu and P. Zanardi, Geometric quantum gates that are robust against stochastic control errors, Phys. Rev. A {\bf 72}, 020301(R) (2005).
\bibitem{local2} P. J. Leek, J. M. Fink, A. Blais, R. Bianchetti, M. G\"{o}ppl, J. M. Gambetta, D. I. Schuster, L. Frunzio, R. J. Schoelkopf, and A. Wallraff, Observation of Berry's Phase in a Solid-State Qubit, Science {\bf 318}, 1889 (2007).
\bibitem{local3} S. Berger, M. Pechal, A. A. Abdumalikov, Jr., C. Eichler, L. Steffen, A. Fedorov, A. Wallraff, and S. Filipp, Exploring the effect of noise on the Berry phase, Phys. Rev. A {\bf 87}, 060303(R) (2013).
\bibitem{GQC1}  J. A. Jones, V. Vedral, A. Ekert, and G. Castagnoli, Geometric quantum computation using nuclear magnetic resonance, Nature (London) {\bf 403}, 869 (2000).
\bibitem{GQC2}  L. M. Duan, J. I. Cirac, and P. Zoller, Geometric Manipulation of Trapped Ions for Quantum Computation, Science {\bf 292}, 1695 (2001).
\bibitem{GQC3} L. A. Wu, P. Zanardi, and D. A. Lidar, Holonomic Quantum Computation in Decoherence-Free Subspaces, Phys. Rev. Lett. {\bf 95}, 130501 (2005).
\bibitem{Berry} M. V. Berry, Quantal phase factors accompanying adiabatic changes, Proc. R. Soc. A {\bf 392}, 45 (1984).
\bibitem{matrix} F. Wilczek and A. Zee, Appearance of Gauge Structure in Simple Dynamical Systems, Phys. Rev. Lett. {\bf 52}, 2111 (1984).
\bibitem{AA} Y. Aharonov and J. Anandan, Phase Change during a Cyclic Quantum Evolution, Phys. Rev. Lett. {\bf 58}, 1593 (1987).
\bibitem{single1} P. Solinas, P. Zanardi, N. Zanghi, and F. Rossi, Nonadiabatic geometrical quantum gates in semiconductor quantum dots, Phys. Rev. A {\bf 67}, 052309 (2003).
\bibitem{single2} Y. Ota and Y. Kondo, Composite pulses in NMR as nonadiabatic geometric quantum gates, Phys. Rev. A {\bf 80}, 024302 (2009).
\bibitem{single3} P. Z. Zhao, X. D. Cui, G. F. Xu, E. Sj\"{o}qvist, and D. M. Tong, Rydberg-atom-based scheme of nonadiabatic geometric quantum computation, Phys. Rev. A {\bf 96}, 052316 (2017).
\bibitem{single4} T. Chen and Z. Y. Xue, Nonadiabatic Geometric Quantum Computation with Parametrically Tunable Coupling, Phys. Rev. Appl. {\bf 10}, 054051 (2018).
\bibitem{single5} K. Z. Li, P. Z. Zhao, and D. M. Tong, Approach to realizing nonadiabatic geometric gates with prescribed evolution paths, Phys. Rev. Research, {\bf 2}, 023295 (2020).
\bibitem{multiple1} S. L. Zhu and Z. D. Wang, Implementation of Universal Quantum Gates Based on Nonadiabatic Geometric Phases, Phys. Rev. Lett. {\bf 89}, 097902 (2002).
\bibitem{multiple2} S. L. Zhu and Z. D. Wang, Universal quantum gates based on a pair of orthogonal cyclic states: Application to NMR systems, Phys. Rev. A {\bf 67}, 022319 (2003).
\bibitem{multiple3}  Y. Ota, Y. Goto, Y. Kondo, and M. Nakahara, Geometric quantum gates in liquid-state NMR based on a cancellation of dynamical phases, Phys. Rev. A {\bf 80}, 052311 (2009).
\bibitem{holonomic1} E. Sj\"{o}qvist, D. M. Tong, L. M. Andersson, B. Hessmo, M. Johansson, and K. Singh, Non-adiabatic holonomic quantum computation, New J. Phys. {\bf 14}, 103035 (2012).
\bibitem{holonomic2} G. F. Xu, J. Zhang, D. M. Tong, E. Sj\"{o}qvist, and L. C. Kwek, Nonadiabatic Holonomic Quantum Computation in Decoherence-Free Subspaces, Phys. Rev. Lett. {\bf 109}, 170501 (2012).
\bibitem{composite} G. F. Xu, P. Z. Zhao, T. H. Xing, E. Sj\"{o}qvist, and D. M. Tong, Composite nonadiabatic holonomic quantum computation, Phys. Rev. A {\bf 95}, 032311 (2017).
\bibitem{NHQC} E. Herterich and E. Sj\"{o}qvist, Single-loop multiple-pulse nonadiabatic holonomic quantum gates, Phys. Rev. A {\bf 94}, 052310 (2016).
\bibitem{NHQC+} B. J. Liu , X. K. Song, Z. Y. Xue, X. Wang, and M. H. Yung, Plug-and-Play Approach to Nonadiabatic Geometric Quantum Gates, Phys. Rev. Lett. {\bf 123}, 100501 (2019).
\bibitem{schemes1} S. L. Zhu and Z. D. Wang, Unconventional Geometric Quantum Computation, Phys. Rev. Lett. {\bf 91}, 187902 (2003).
\bibitem{schemes2} S. B. Zheng, Unconventional geometric quantum phase gates with a cavity QED system, Phys. Rev. A {\bf 70}, 052320 (2004).
\bibitem{schemes3} P. Z. Zhao, G. F. Xu, and D. M. Tong, Nonadiabatic geometric quantum computation in decoherence-free subspaces based on unconventional geometric phases, Phys. Rev. A {\bf 94}, 062327 (2016).
\bibitem{circuits1} Y. Xu, W. Cai, Y. Ma, X. Mu, L. Hu, T. Chen, H. Wang, Y. P. Song, Z. Y. Xue, Z. Q. Yin \emph{et al.}, Single-Loop Realization of Arbitrary Nonadiabatic Holonomic Single-Qubit Quantum
Gates in a Superconducting Circuit, Phys. Rev. Lett. {\bf 121}, 110501 (2018).
\bibitem{circuits2} Z. X. Zhang, P. Z. Zhao, T. H. Wang, L. Xiang, Z. L. Jia, P. Duan, D. M. Tong, Y. Yin, and G. P. Guo, Single-shot realization of nonadiabatic holonomic gates with a
superconducting Xmon qutrit, New J. Phys. {\bf 21}, 073024 (2019).
\bibitem{circuits3} Y. Xu, Z. Hua, T. Chen, X. Pan, X. Li, J. Han, W. Cai, Y. Ma, H. Wang, Y. P. Song \emph{et al.}, Experimental Implementation of Universal Nonadiabatic Geometric Quantum Gates
in a Superconducting Circuit, Phys. Rev. Lett. {\bf 124}, 230503 (2020).
\bibitem{resonance1} J. F. Du, P. Zou, and Z. D. Wang, Experimental implementation of high-fidelity unconventional geometric quantum gates
using an NMR interferometer, Phys. Rev. A {\bf 74}, 020302(R) (2006).
\bibitem{resonance2} H. Li, Y. Liu, and G. L. Long, Experimental realization of single-shot nonadiabatic holonomic gates in nuclear spins, Sci. China. Phys. Mech. Astron. {\bf 60}, 080311 (2017).
\bibitem{ions} D. Leibfried, B. DeMarco, V. Meyer, D. Lucas, M. Barrett, J. Britton, W. M. Itano, B. Jelenkovi\'{c}, C. Langer, T. Rosenband \emph{et al.}, Experimental demonstration of a
robust, high-fidelity geometric two ion-qubit phase gate, Nature {\bf 422}, 412 (2003).
\bibitem{diamond1} A. C. Silvia, L. Andrii, W. H. Stefan, and B. Gopalakrishnan, Room temperature high-fidelity holonomic single-qubit gate on a solid-state spin, Nat Commun {\bf 5}, 4870 (2014).
\bibitem{diamond2} C. Zu, W. B. Wang, L. He, W. G. Zhang, C. Y. Dai, F. Wang, and L. M. Duan, Experimental realization of universal geometric quantum gates with solid state spins, Nature (London) {\bf 514}, 72 (2014).
\bibitem{diamond3} B. B. Zhou, P. C. Jerger, V. O. Shkolnikov, F. J. Heremans, G. Burkard, and D. D. Awschalom, Holonomic Quantum Control by Coherent Optical Excitation in Diamond, Phys. Rev. Lett. {\bf 119}, 140503 (2017).
\bibitem{Complex}  On the complex Bloch sphere, both any quantum state $|\phi_{+}\rangle=\cos{\frac{\tilde{\theta}}{2}}|+\rangle+\sin{\frac{\tilde{\theta}}{2}}e^{ i\tilde{\varphi}}|-\rangle$ and its corresponding complex vector $\tilde{\bm{n}}=(\sin\tilde{\theta}\cos\tilde{\varphi},\sin\tilde{\theta}\sin\tilde{\varphi},\cos\tilde{\theta})$ maintain a unit modulus by redefining the inner product as
$\langle\phi_{+}|\phi_{+}\rangle_{B}=\langle\tilde{\phi}_{+}|\phi_{+}\rangle$ and $(\tilde{\bm{n}},\tilde{\bm{n}})_{B}=\tilde{\bm{n}}\cdot\tilde{\bm{n}}$, respectively. Here $|\tilde{\phi}_{+}\rangle=\cos{\frac{\tilde{\theta}^{*}}{2}}|+\rangle+\sin{\frac{\tilde{\theta}^{*}}{2}}e^{ i\tilde{\varphi}^{*}}|-\rangle$


\bibitem{optical1} S. Klaiman, U. G\"{u}nther, and N. Moiseyev, Visualization of Branch Points in PT-Symmetric Waveguides, Phys. Rev. Lett. {\bf 101}, 080402 (2008).
\bibitem{optical2} C. E. R\"{u}ter, K. G. Makris, R. E. Ganainy, D. N. Christodoulides, M. Segev, and D. Kip, Observation of parity-time symmetry in optics, Nature Phys. {\bf 6} 192-195 (2010).
\bibitem{decay} W. E. Lamb, R. R. Schlicher, and M. O. Scully, Matter-field interaction in atomic physics and quantum optics, Phys. Rev. A {\bf 36}, 2763 (1987).

\bibitem{SM} Supplemental Material for the details of ¡°complex Bloch sphere¡± and ¡°complex A-A connection¡±.
\bibitem{space} P. J. Zhao, W. Li, H. Cao, S. W. Yao, and L. X. Cen, Exotic dynamical evolution in a secant-pulse-driven quantum system, Phys. Rev. A {\bf 98}, 022136 (2018).
\bibitem{azimuthal} Following the expression of Eq. (\ref{externals}), the imperfect control of the pulse parameters represented as $\Omega_{x,y,z}'(t)$ could not be explicitly expressed by pulse parameters $\Omega_{x,y,z}(t)$ and the deviation of the azimuthal angle $\tilde{\theta}(t)$.
\bibitem{fidelity} W. Li, Invariant-based inverse engineering for fast nonadiabatic geometric quantum computation, New J. Phys. {\bf 23}, 073039 (2021).
\bibitem{dynamical} S. B. Zheng, C. P. Yang, and F. Nori, Comparison of the sensitivity to systematic errors between nonadiabatic non-Abelian geometric gates and their dynamical counterparts, Phys. Rev. A {\bf 93} 032313 (2016).
\bibitem{system} D. Jaksch, J. I. Cirac, P. Zoller, S. L. Rolston, R. C\^{o}t\'{e}, and M. D. Lukin, Fast Quantum Gates for Neutral Atoms, Phys. Rev. Lett. {\bf 85}, 2208 (2000).
\end{thebibliography}
\end{document}